\documentclass[manuscript]{aastex}

\slugcomment{To be published in ApJ Letters}

\shorttitle{Young Planetary-Mass Object in $\rho$ Oph}
\shortauthors{Marsh et al.}

\begin{document}

\title{A Young Planetary-Mass Object in the $\rho$ Oph Cloud Core}


\author{Kenneth A. Marsh, J. Davy Kirkpatrick and Peter Plavchan}
\affil{Infrared Processing and Analysis Center, California Institute of 
Technology 100-22, Pasadena, CA 91125;
\email{kam@ipac.caltech.edu, davy@ipac.caltech.edu, plavchan@ipac.caltech.edu}}

\begin{abstract}
We report the discovery of a young planetary-mass brown dwarf 
in the $\rho$ Oph cloud core. The object was identified as such with the aid of
a 1.5--2.4 $\mu$m low-resolution spectrum obtained using the NIRC 
instrument on the Keck I telescope.  Based on the COND model, the observed spectrum is
consistent with a reddened ($A_V\sim15-16$) brown dwarf whose effective 
temperature is in the range 1200--1800 K.  For an assumed age 
of 1 Myr, comparison with isochrones further constrains the temperature 
to $\sim1400$ K and suggests a mass of $\sim2$--3 Jupiter masses.  
The inferred temperature is suggestive of an early T spectral type, which
is supported by spectral morphology consistent with weak methane absorption. 
Based on its inferred distance ($\sim100$ pc) and the presence of 
overlying visual absorption,
it is very likely to be a $\rho$ Oph cluster member. In addition, given
the estimated spectral type, it may be
the youngest and least massive T dwarf found so far.  Its
existence suggests that the initial mass function for the $\rho$ Oph
star-forming region extends well into the planetary-mass regime.
\end{abstract}


\section{Introduction}

Free-floating planetary-mass objects have been purported in
young clusters such as $\sigma$ Orionis \citep{zap02} and the Trapezium
\citep{lucas06}.
These are brown dwarfs whose masses are below the deuterium-burning
limit of 13 Jupiter masses ($M_{\rm Jup}$), and their existence imposes severe 
restrictions on brown dwarf formation models.  The models must account both 
for the generation of low-mass cloud fragments and also for their 
subsequent collapse.  In principle, the required masses can be produced by 
mechanisms such as fragmentation of collapsing protostellar cores 
\citep{boss01} or encounters between protostellar disks \citep{shen06},
but this still leaves the problem of providing
the necessary cooling to allow gravity to overcome gas pressure.
In this regard, recent work by \citet{whit06} suggests
that masses as low as 1-4 $M_{\rm Jup}$ are possible, the lower limit
being independent of the particular star formation scenario.
However, difficulties still arise when subsequent accretion is taken into
account, since this can greatly increase the final mass of the object.  
Subsequent dynamical ejection may be necessary in order to prevent 
this mass gain (see, for example, \citet{reip01}).
A key constraint on these models could be provided by observational
determination of the form of the initial mass function, and, in particular, 
whether there is a minimum mass for brown dwarf formation.  The latter
would provide a key test of the prediction of \cite{whit06}, even though
additional tests would be required in order to distinguish between the possible 
fragmentation mechanisms.

With this in mind, we have begun a spectroscopic study of objects in the
$\rho$ Oph cloud core whose near-infrared colors are suggestive of
low-mass brown dwarfs.  This region is a particularly suitable site for the 
study of young brown dwarfs due to its youth, high rate of low-mass star 
formation and its relative proximity at a distance of $\sim120$--130 pc 
\citep{loin08,mama08,lom08}.
We have produced a list of brown dwarf candidates in a $1^\circ\times 9'$
region of the cloud core, based on deep integration $J$, $H$, and $K_s$
images from the stacked calibration scans of 2MASS \citep{cutri06} together 
with {\em Spitzer\/} IRAC images at 3.6, 4.5, 5.8 and 8.0 $\mu$m.  
Our selection technique is based on the estimated effective temperature, 
obtained from  fits of the observed spectral energy
distributions (SEDs) to atmospheric models (Marsh et al., in preparation).
In a pilot program of spectroscopic observations, we have observed seven 
candidates from
this list using the NIRC instument on the Keck I telescope, and
now present the results.  In particular, we report the first discovery of a
planetary-mass brown dwarf in $\rho$ Oph.

\section{Observations and Data Reduction}

We selected seven brown dwarf candidates whose near-infrared
colors suggested effective temperatures less than 2000 K.  These objects are
listed in Table \ref{tbl-1}; each is labeled with an ID number 
based on the ordering of our source extractions in the 
$1^\circ\times9'$ 2MASS Deep Field; the full list will be presented
in the near future by Marsh et al.  We observed these objects
using the NIRC instrument \citep{mat94} on the
10-m Keck I telescope on the night of 2009 May 27 (UT).  NIRC was used
in grism mode (grism gr120), with a slit width of 3.5 pixels, to produce
low-resolution ($R\sim90$) spectra in the wavelength range 1.5--2.4 $\mu$m.
Conditions were essentially clear except for 
light cirrus haze, and the seeing was $\sim0.6''$. 

Targets were located on the slit by first imaging them in the $K$-band
continuum. They were then observed in grism mode in
sets of 30 s or 60 s exposures, using a nodding pattern in which the
telescope was dithered at intervals of
$5''$ along the slit.  Typically, 5 such sets were coadded.

The coadded images were pair-wise subtracted to remove the dark current and
sky background, and divided by a flat-field image to remove pixel-to-pixel
gain variations. The latter image was derived from observations of the telescope
dome after subtraction
of the dark response.  It was also used in the construction of a bad pixel
mask. Correction for the
spectral response of the instrument and absorption by the telluric
bands of the atmosphere was then obtained from observations of A0 stars
Oph S1 and HD 161743. The correction was
applied by multiplying the spectrum of a program star by a 10,000 K
black body spectrum and dividing by the observed spectrum of the A0
star.  In doing so, it was necessary to deredden the Oph S1 spectrum 
to correct for
the known 10 magnitudes of visual absorption \citep{gag04}.  For telluric
correction we also utilized an observation of NGC 4361, a planetary nebula
whose $HK$ spectrum is dominated by the continuum of the hot central star.
Wavelength calibration was accomplished using a cubic fit to a set of
atmospheric OH lines identified in sky observations based on
\citet{rous2000}, and a set of
nebular lines from an observation of another planetary nebula, G049.3+88.1.

\begin{figure}
\epsscale{1.0}
\plotone{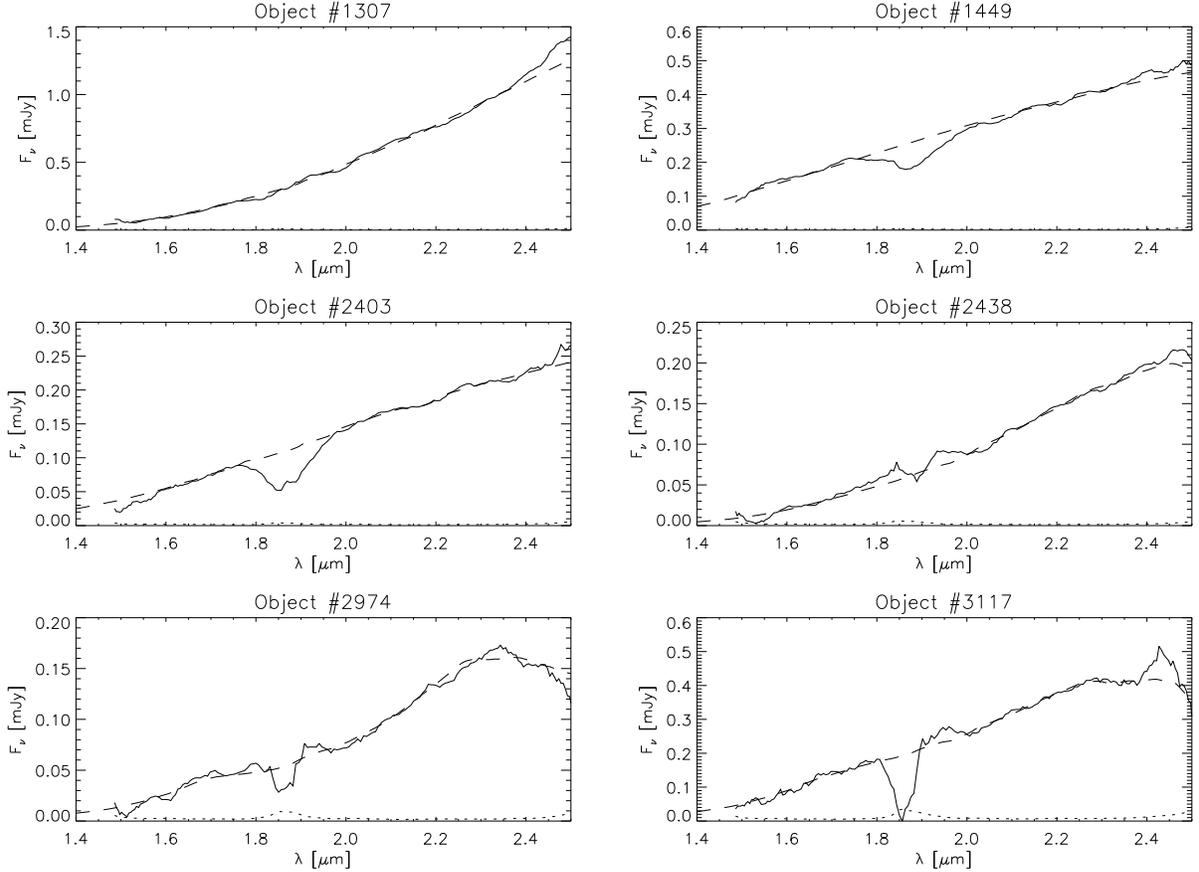}
\caption{Spectra of six of the seven candidate objects.
In each case, the solid
line represents the observed spectrum and the dotted line represents
the uncertainty due to uncorrelated measurement noise; the dashed line 
represents the best-fit model spectrum.  A boxcar averaging window of width
9 pixels (0.045 $\mu$m) has been applied in all cases.
Peak $S/N$ values for the six spectra (after smoothing), in order of object 
ID number, are: 248, 192, 156, 126, 87, and 57.
Note that the pronounced dips
at $\sim1.85$ $\mu$m represent an artifact due to imperfect correction for
the deep telluric absorption bands.}
\label{fig1}
\end{figure}

\begin{figure}
\epsscale{1.0}
\plotone{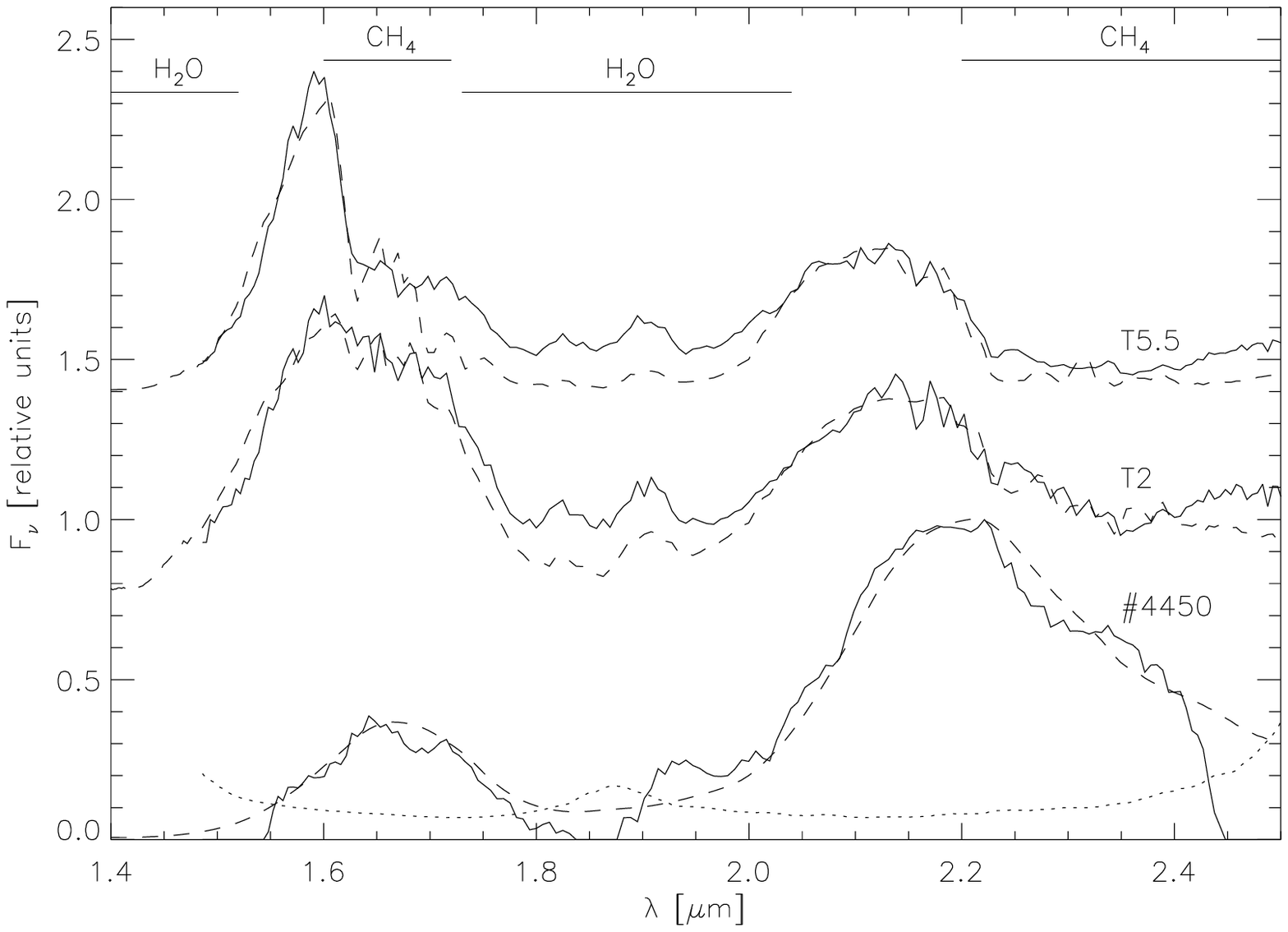}
\caption{Spectra of object \#4450 and the two known T dwarfs,
SDSS 1254-0122 (T2) and 2MASS 1503+2525 (T5.5), normalized to peak fluxes
of 0.075, 3.6, and 6.2 mJy, respectively.  
For clarity, the T2 and T5.5
spectra have been displaced vertically by 0.7 and 1.4 units, respectively.
In each case, the solid
line represents the observed spectrum and the dashed line
represents the best-fit model spectrum. The dotted line represents the RMS
measurement noise in the \#4450 spectrum. A boxcar averaging window of width
21 pixels (0.1 $\mu$m) was used in the estimation of the \#4450 spectrum,
but no averaging was used for the other two objects.  
The peak SNR of the smoothed \#4450 spectrum was 13.5;  the peak SNRs
for SDS 1254-0122 and 2MASS 1503+2525 were 1188 and 1848, respectively.
The approximate extents
of the water and methane absorption bands are indicated at the top.}
\label{fig2}
\end{figure}

The maximum likelihood estimate of the spectrum of each program object was obtained
using all available observations for that object simultaneously. 
This necessitated spatial co-registration 
of the individual subtracted nods, accomplished by adjusting the image
offset parallel to the slit based on maximum correlation of the signal
with the wavelength-dependent spatial point spread function (PSF).  The latter
was estimated using observations of strong calibrators.  The program star 
spectrum was then obtained from a series of maximum likelihood flux estimates
as a function of wavelength using all observations which fell within a
rectangular ``fitting" window 
of width typically 9 pixels in both the spatial and spectral directions,
where 1 pixel corresponds to $0.15''$ and $\sim0.005$ $\mu$m, respectively;
the spectral boxcar width was, however, increased to 21 pixels for 
\#4450, the object with lowest $S/N$.  The estimation procedure involved
knowledge of the measurement noise, estimated from adjacent strips
on the image, external to the strip containing the signal.  In order to
minimize the effect of spikes due to cosmic ray hits, it was necessary to
use trimmed averaging during noise estimation, and $3\sigma$ outlier rejection
during spectral estimation.  The estimated spectra
of our seven program objects
are shown by the solid lines in Figures \ref{fig1} and \ref{fig2}.  
Also included in Figure \ref{fig2}, for comparison,
are our spectra of two known T dwarfs, 
SDSS 1254-0122 (T2) and 2MASS 1503+2525 (T5.5).

\section{Model Fitting}

The observations have been fitted to a grid of synthetic spectra 
based on photospheric models described by \citet{all01,all03} and \citet{haus99},
and downloaded from the Lyon group's
website\footnote{\url{http://phoenix.ens-lyon.fr/Grids}}.
The grid covers the range $T_{\rm eff}=100$--10,000 K in effective temperature
and $\log g=2.5$--6.0 cm s$^{-2}$ in surface gravity; we  have assumed solar 
metallicity.  The temperature range is spanned by
four models (COND, SETTL, DUSTY and NextGen), each of which covers a particular
regime with respect to dust grain formation; 
the COND model is applicable to
methane dwarfs ($T\stackrel{<}{_\sim}1500$ K). 

For each of these models we have calculated the inverse-variance weighted 
sum of squares of residuals between the model (smoothed to the resolution
of the observations) and the observed spectrum
of a given object. During this procedure
it was necessary to redden the model spectra
since objects in the $\rho$ Oph cloud are seen through
a substantial amount of extinction; for this purpose the \citet{car89}
reddening law was used.
For each object, the unknowns were
therefore:  $T_{\rm eff}$, $\log g$, $A_V$, and the model type. 
Maximum likelihood estimates of these parameters were obtained by
minimizing the mean square residual over the wavelength range 1.5--2.4 $\mu$m
(excluding 1.7--2.0 $\mu$m to avoid the deep telluric absorption bands), and 
the results are presented in the last four columns of Table \ref{tbl-1}. 
The corresponding model spectra are plotted as dashed lines in 
Figures \ref{fig1} and \ref{fig2}.

One of the objects (\#4450) has a spectrum which resembles a reddened version
of SDSS 1254-0122 (T2), and the model fitting results confirm its identity as
a moderately cool brown dwarf.  The spectrum is suggestive of a low 
gravity object, as evidenced
by the steeper falloff on the short wavelength side of the $H$-band peak 
with respect to that of the field dwarfs in Figure \ref{fig2}, and the 
displacement of the $K$-band peak to a longer wavelength---both of these 
features are consistent with the much deeper H$_2$O absorption expected in a 
lower gravity object.

\begin{figure}
\epsscale{0.8}
\plotone{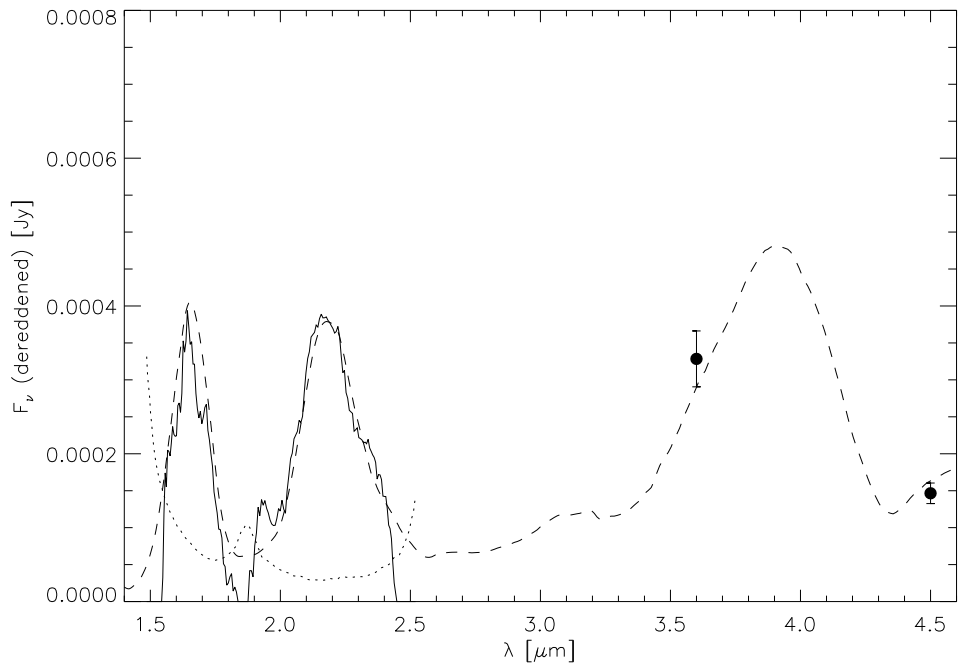}
\caption{De-reddened spectrum of object \#4450 (solid line), and
the best-fit model (dashed line) based on
$T_{\rm eff}=1400$ K and $\log g=3.0$.
The dotted line represents the uncertainty in the observed spectrum and
the filled circles represent the IRAC fluxes at 3.6 and 4.5 $\mu$m.}
\label{fig3}
\end{figure}

The maximum likelihood parameter
estimates based on the observed NIRC spectrum are: $T_{\rm eff}=1600\,^{+250}
_{-450}$ K, $\log g=4.5\,^{+0.8}_{-2.0}$ and $A_V=24.5$ with an uncertainty
of $\sim\pm10$ mag.  The large uncertainty in $A_V$ is due to the relatively
low $S/N$ which allows for a fairly large range in model temperatures. 
To gain more leverage for the $A_V$ estimate and to further constrain
the other parameters, we repeated the model fitting incorporating the
IRAC [3.6] and [4.5] magnitudes ($15.77\pm0.13$ and $15.85\pm0.10$,
respectively; Marsh et al., in preparation).  We obtained: 
$T_{\rm eff}=1400\pm{100}$ K, $\log g=3.0^{+1.7}_{-0.5}$, $A_V=15.6\pm2.5$;
the corresponding fit to the dereddened spectrum (+ IRAC fluxes) is 
shown in Figure \ref{fig3}.

\section{Discussion}

The results of model fitting indicate that:

\begin{itemize}
 \item[1.] The estimated temperatures for the two known T dwarfs are consistent 
with the previously-published spectral types \citep{burg03,dahn02}.  They
also indicate surface gravities characteristic of mature brown dwarfs, and
zero visual absorption within the error bars, as expected.  These two results 
have therefore provided a useful check on our observing and data reduction 
techniques. This is particularly significant in view of
\citet{burg04}'s finding that model fits to some T dwarf spectra yield
anomalously low surface gravities.  However that bias occurred only
for late T dwarfs (T7 onwards), and hence there is no disagreement with
the present results.
 \item[2.] For all except object \#4450, the spectroscopically-estimated temperature is
substantially higher than the $T_{\rm eff}<2000$ K values inferred from our
previous SED fits.  These other six objects are most likely T Tau stars and/or
reddened background stars. One possible exception is \#2974, whose
error bars place it within reach of being a brown dwarf near the 
hydrogen-burning limit (its observed spectrum is consistent with 
$T_{\rm eff}=2800$ K \& $\log g=4.0$, corresponding to a mass as low as
0.05 $M_\odot$ based on the isochrones of \citet{bar03}). 
In our forthcoming paper (Marsh et al., in preparation)
we discuss the likely reason for the six underestimated temperatures, and have 
used the results to improve our SED fitting procedure. 
 \item[3.] Object \#4450 is a low-gravity brown dwarf whose estimated
temperature suggests an early T spectral type.  This is supported by the 
presence of weak, broad absorption features on the long wavelength sides of 
the $H$ and $K$ peaks (see Figure \ref{fig2}), suggestive of methane.
Definitive identification of methane will, however, need to await future
observations at higher spectral resolution and signal to noise ratio.
\end{itemize}

\begin{figure}
\epsscale{0.8}
\plotone{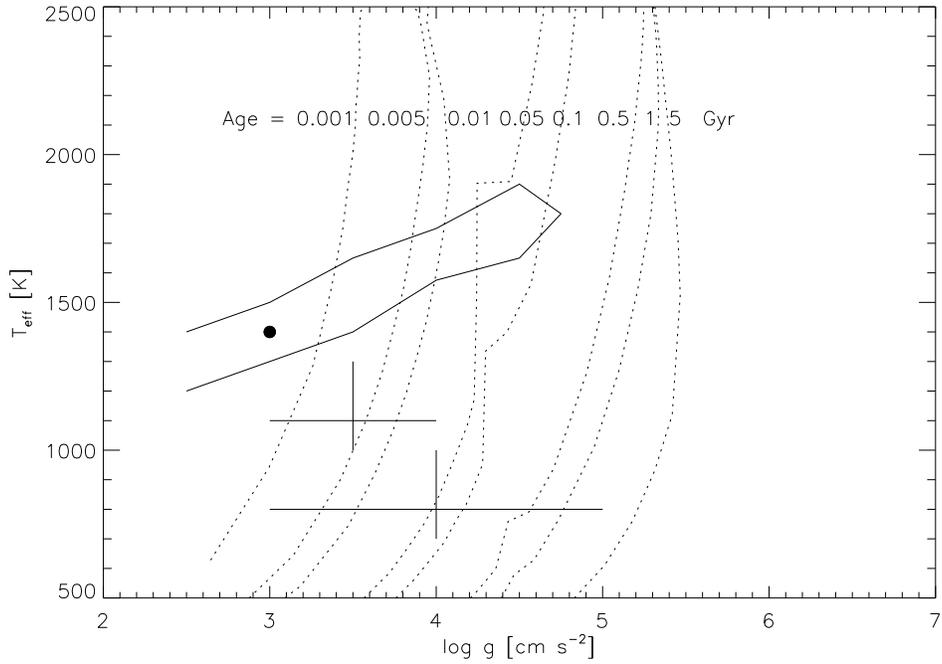}
\caption{Results of model fitting for object \#4450, in the 
$\log g$--$T_{\rm eff}$ plane.  The best fit to
the observed spectrum is indicated by the filled circle, and the $1\sigma$
uncertainty is indicated by the solid contour.  The dotted lines represent
isochrones from \citet{bar03}. For comparison, the crosses represent the
estimated parameters of S Ori 70, another arguably young object of 
comparable mass \citep{zap02,martin03}.
}
\label{fig4}
\end{figure}

Figure \ref{fig4} shows the estimated location of \#4450 in the 
$\log g$--$T_{\rm eff}$ plane, together with the $1\sigma$ contour of the
model fitting solution.  Also superposed on the figure is a set of isochrones
for the COND model, from \citet{bar03}.  If we take the estimated age of 1 Myr for
the $\rho$ Oph cloud \citep{luh99,prat03,wil05}, the isochrones further
constrain the parameters of this object. Specifically, the intersection of
the $1\sigma$ contour and the 1 Myr isochrone corresponds to $T_{\rm eff}\sim
1400$ K and $\log g\sim3.3$; such a temperature reinforces our estimate of
early T spectral type.  The corresponding mass would be $\sim2-3$ 
$M_{\rm Jup}$, making \#4450 a candidate for the lowest-mass object outside our
solar system to have been imaged to date.  
Another candidate for this distinction
is S Ori 70 \citep{zap02,martin03}, whose parameters are also indicated on
Figure \ref{fig4}.  We note, however, that the identification of S Ori 70
as a young, low-gravity member of the $\sigma$ Orionis cluster has been called
into question by \citet{burg04}, who find that the observed spectra of that
object could equally well be interpreted in terms of an old field T dwarf of 
much higher mass.

We can estimate a distance to \#4450 by comparing the apparent 
and absolute magnitudes, obtained from the observations and model, respectively.
If we assume an age of 1 Myr, then the corresponding isochrone in 
Figure \ref{fig4} indicates a range 1360--1610 K of $T_{\rm eff}$ values
from the spectral fits; the corresponding range in $M_K$, obtained
by spline interpolation of the isochronal values from \citet{bar03}, is then
11.64--10.72.  Comparison with the dereddened apparent $K$ magnitude of 
$15.86\pm0.23$, assuming that all of the uncertainties add
in quadrature, then leads to a $\pm1\sigma$ range in distance of 68--111 pc.
Although this might suggest that \#4450 is closer than $\rho$ Oph,
based on most distance estimates (120-130 pc), this range is 
in fact, consistent with the recent $119\pm6$ pc estimate of \citet{lom08} 
when account is taken of the finite thickness of the cloud along the line of 
sight ($28^{+29}_{-19}$ pc).  Our results are therefore
consistent with \#4450 being located near the front edge of the
$\rho$ Oph cloud. If the object is older than 1 Myr, the estimated distance
decreases, since older objects are less bloated and therefore must be closer
to maintain the same flux.  If, for example, we assume an age of 100 Myr
(the largest age within the $1\sigma$ contour of Figure \ref{fig4}),
the distance upper limit gets reduced to $\sim86$ pc.

What evidence do we have that object \#4450 is a $\rho$ Oph cluster member and
not simply a foreground or background T dwarf?  We can readily exclude the
former possibility because of the presence of substantial visual extinction.
Also, an old background T dwarf is excluded since it would have insufficient 
flux, based on the discussion in the previous paragraph.  However, since the
line of sight to \#4450 intersects the southeastern portion of the 
Upper Sco association \citep{dez99}, we must consider
the latter as an another possible location, which would imply
an age of $\sim5$ Myr \citep{mama08}.  However, the 145 pc distance
of Upper Sco \citep{mama08} is almost $3\sigma$ beyond the estimated
distance of \#4450 and hence membership in $\rho$ Oph is much more likely.
In either case, the estimated mass would still be 
$\stackrel{<}{_\sim}3$ $M_{\rm Jup}$.

Our conclusions regarding the nature of this object could be tested
by future observations at higher spectral resolution ($\sim1000$), with
an instrument such as NIRSPEC \citep{mcl98},
particularly if the wavelength range were extended to include the $J$-band. 
Measurement of key flux ratios of H$_2$O and CH$_4$ features 
(see \citet{mcl03}) would then facilitate more accurate spectral typing,
and the gravity estimate could be refined using the K I doublet lines near 
1.25 $\mu$m.  Comparison of the resulting $T_{\rm eff}$ and $\log g$ 
with the COND isochrones might then lead to a better constraint on the
age which, in the present analysis, has been assumed to be 1 Myr based on 
sky location in the $\rho$ Oph cloud.

Our detection of this object suggests that the initial mass function for the 
$\rho$ Oph star-forming region extends well into the planetary-mass regime.
Object \#4450 is, however, only one of about a dozen objects within our 
$1^\circ\times9'$ region of study whose near-infrared SEDs are similarly 
indicative of planetary-mass brown dwarfs, and we plan to obtain near 
infrared spectra of these objects in the near future.  We thereby hope
to accumulate some statistics on the low end of the brown dwarf mass function
to further our goal of constraining the formation models. In this regard, 
we also look forward to the expected wealth of information from the Wide-field 
Infrared Survey Explorer (WISE) mission, which will provide an inventory of 
low-mass cold brown dwarfs in the immediate solar neighborhood.

\acknowledgments 

We thank the referee (A. Muench) for helpful comments.
The data presented herein were obtained at the W.M. Keck Observatory, which is operated as a scientific partnership among the California Institute of Technology, the University of California and the National Aeronautics and Space Administration. The Observatory was made possible by the generous financial support of the W.M. Keck Foundation. The authors wish to recognize and acknowledge the very significant cultural role and reverence that the summit of Mauna Kea has always had within the indigenous Hawaiian community.  We are most fortunate to have the opportunity to conduct observations from this mountain. Our research also utilized 
the Simbad database, operated at CDS, Strasbourg, France.
The work was carried out at IPAC/Caltech and was supported by a grant from 
the NASA Astrophysics Data Analysis Program and by a Keck PI data award.

\clearpage

\begin{deluxetable}{ccccccccccc}
\tabletypesize{\scriptsize}
\tablecaption{Observational and model-fitted parameters. \label{tbl-1}}
\tablewidth{0pt}
\tablehead{
\colhead{Object ID} & 
\colhead{Type\tablenotemark{a}} &
\colhead{RA} & \colhead{Dec} &
\colhead{$K$} & 
\colhead{$t_{\rm int}$\tablenotemark{b}} &
\colhead{A.M.\tablenotemark{c}} & 
\colhead{$T_{\rm eff}$} &
\colhead{$\log g$\tablenotemark{d}} &
\colhead{$A_V$\tablenotemark{e}} & 
\colhead{Mod.\tablenotemark{f}} \\  
& & &  & & \colhead{[s]} & & \colhead{[K]} & \colhead{[cm s$^{-2}$]} & 
\colhead{[mag]} & 
}
\startdata
\#1307\tablenotemark{g} & bdc & 16 27 32.89 &  $-24$ 28 11.4 &    14.92 
& 360 &   1.68 & $>4800$ & indet. & 36.1 & N \\
\#1449 & bdc & 16 27 30.36 &  $-24$ 20 52.2 &    15.67 &  900 &   1.40 
& $>5700$ & indet. & 21.4 & N \\
\#2403 & bdc & 16 27 21.63 &  $-24$ 32 19.2 &    16.42 &  900 &   1.70 
& $>5200$ & indet. & 24.5 & N \\
\#2438 & bdc & 16 27 09.37 &  $-24$ 32 14.9 &    16.74 &  900 &   1.92 
& $>3500$ & indet. & 32.7 & D \\
\#2974 & bdc & 16 27 16.74 &  $-24$ 25 39.0 &    16.86 &  900 &   2.20
& $3200^{+300}_{-400}$ & indet. & 25.9 & D \\
\#3117 & bdc & 16 27 17.68 &  $-24$ 25 53.5 &    15.70 &  900 &   2.88 
& $3800^{+200}_{-400}$ & indet. & 24.7 & D \\
\#4450 & bdc & 16 27 25.35 &  $-24$ 25 37.5 &    17.71 &  1080 &   1.43 
& 1400$\pm$100 & $3.0^{+1.7}_{-0.5}$ & 15.6 & C \\
SDSS 1254-0122 & T2 & 12 54 53.90 & $-01$ 22 47.4  & 13.84 & 450 & 1.06
& 1500$\pm$100 & 5.0$\pm$0.5 & 0.0 & C \\
2MASS 1503+2525 & T5.5 & 15 03 19.61 &  $+25$ 25 19.6 & 13.96 & 900 & 1.06 
& 1100$\pm$100 & 5.0$\pm$0.5 & 0.7 & S \\
NGC 4361\tablenotemark{h} & PN$\star$ & 12 24 30.76 & $-18$ 47 05.4 & 14.02 
& 450 &1.29& \nodata & \nodata & \nodata & \nodata \\
Oph S1\tablenotemark{h} & A0 & 16 26 34.17 & $-24$ 23 28.3 & 6.32 & 6.1 
& $1.74,\!1.80$ & \nodata & \nodata & \nodata & \nodata \\
HD 161743\tablenotemark{h} & A0 & 17 48 57.93 & $-38$ 07 07.5 & 7.57 & 6.1 
& 3.09 &  \nodata & \nodata & \nodata & \nodata \\
\enddata
\tablecomments{Spectral types and $K$ magnitudes for the two known 
T dwarfs are from \citet{burg03}, \citet{leg00}, and \citet{dahn02}. 
The last four columns of the table represent the results of model
fitting, as described in the text.}
\tablenotetext{a}{Designation ``bdc" means that the object was a brown dwarf candidate}
\tablenotetext{b}{Integration time}
\tablenotetext{c}{Air mass of observation}
\tablenotetext{d}{Designation ``indet." means that the quantity was 
indeterminate from the model fits}
\tablenotetext{e}{Uncertainty in $A_V$ is $\pm2.5$ mag for \#4450, and $\pm 1.0$
mag in all other cases} 
\tablenotetext{f}{Best fit model: N=NextGen, D=DUSTY, S=SETTL, C=COND}
\tablenotetext{g}{Corresponds to 16273288-2428116 in the 2MASS Point Source Catalog}
\tablenotetext{h}{Used as calibrator}
\end{deluxetable}

\end{document}